\documentclass[aps,pra,twocolumn,superscriptaddress]{revtex4}
\usepackage{amssymb}

\usepackage{graphicx}

\begin{document}

\title{Electromagnetically induced transparency controlled by
a microwave field
}

\author{Hebin Li}
\affiliation{
    Department of Physics and Institute for Quantum Studies,
    Texas A\&M University,
    College Station, Texas 77843-4242
}

\author{Vladimir A. Sautenkov}
\affiliation{
    Department of Physics and Institute for Quantum Studies,
    Texas A\&M University,
    College Station, Texas 77843-4242
}

\affiliation{
   Lebedev Institute of Physics, Moscow 119991, Russia
}

\author{Yuri V. Rostovtsev}
\affiliation{
    Department of Physics and Institute for Quantum Studies,
    Texas A\&M University,
    College Station, Texas 77843-4242
}

\author{George R. Welch}
\affiliation{
    Department of Physics and Institute for Quantum Studies,
    Texas A\&M University,
    College Station, Texas 77843-4242
}

\author{Philip R. Hemmer}
\affiliation{
    Electrical Engineering Department,
    Texas A\&M University,
    College Station, Texas 77843
    }

\author{Marlan\ O.\ Scully}
\affiliation{
    Department of Physics and Institute for Quantum Studies,
    Texas A\&M University,
    College Station, Texas 77843-4242
}

\affiliation{ Applied Physics and Materials Science Group,
Engineering Quad, Princeton University, Princeton, New Jersey 08544
}

\date{\today}

\begin{abstract}
We have experimentally studied propagation of two optical fields in
a dense rubidium (Rb) vapor in the case when an additional microwave
field is coupled to the hyperfine levels of Rb atoms. The Rb energy
levels form a close-lambda three-level system coupled to the optical
fields and the microwave field. It has been found that the maximum
transmission of a probe field depends on the relative phase between
the optical and the microwave fields. We have observed both
constructive and destructive interference in electromagnetically
induced transparency (EIT). A simple theoretical model and a
numerical simulation have been developed to explain the observed
experimental results.
\end{abstract}




\maketitle
\section{Introduction}

Electromagnetically induced transparency (EIT) is based on quantum coherence
\cite{arimondo96,book,harris97pt,marangos98jmo,eit05rmp}
that has been shown to result in many counter-intuitive phenomena. The
scattering via a gradient force in gases~\cite{harris-prl},
the forward Brillouin scattering in ultra-dispersive resonant
media~\cite{matsko00prl,matsko01prl,rost01os},
controlled coherent multi-wave mixing~\cite{rost06prl,rost07ieee},
EIT and slow light in various
media \cite{Hau1999,Kash1999,Budker1999,delay,coussement02prl,matsko,hemmer},
Doppler broadening elimination~\cite{ye},
light induced chirality in a nonchiral medium~\cite{sau05prl},
a new class of entanglement amplifier~\cite{cel} based on correlated
spontaneous emission lasers~\cite{Scully,schleich}
and the coherent Raman scattering enhancement via maximal
coherence in atoms~\cite{harris,merriam} and
biomolecules~\cite{scully02pnas,ZY04jmo,gb05oc,pestov07science}
are a few examples that demonstrate the importance of quantum coherence.

Usually, the EIT has been observed in atoms that have a three-level
configuration such as $\Lambda$, V, and Ladder
schemes~\cite{arimondo96,book}. For these schemes
several theoretical approaches have been developed to provide clear physical
insights. For example, the EIT can be understood in the bare state basis
using quantum coherence, or in the dressed state basis involving
Fano interference, or using the so-called dark and bright
states~\cite{arimondo96,book}.

Natural generalization of the three-level schemes is
the so-called double-$\Lambda$, double-V, double-Ladder, and $\Lambda$-V
schemes \cite{double-lambda}, where two relatively strong optical fields
applied to the atomic system to create coherence, and then a probe field
propagates through the gas of such atoms together with an additional strong
drive field. The probe propagation depends on the parameters of the mediam and
the fields preparing coherence. On the other hand, let us note that the effect
of these two optical drive fields is equivalent to an effective microwave
driving field applied to the system. Furthermore, in some regard, the schemes
that involve two optical fields and a microwave field can be related to the
double-$\Lambda$ scheme.

The systems involving interaction with two optical fields and a low
frequency microwave field coupled to the hyperfine levels
have been in a focus of recent studies~\cite{mu-wave-eit}.
For example, microwave interaction~\cite{Hemmer1990} has been used to excite
the Raman trapped state and it was shown that there is influence of the
microwave field on the CPT in a $\Lambda$ system; four-wave mixing(FWM) of
optical and microwave fields has been demonstrated~\cite{Zibrov2002} in
Rb vapor. A microwave field has also been used to study double dark
resonances~\cite{Yelin2003}.
It has been shown~\cite{Yamamoto1998} that, in a V-scheme three-level system of
Pr$^{3+}$:YAlO$_3$ that was excited by a microwave driving field
and two optical probe fields, the probe transmission was either
constructively or destructively affected by the phase of the
microwave field.

Recently, the phase effects in EIT systems has been studied
\cite{pack06pra,frank07jmo,sautenkov08jmo}, where the transient
times of the refractive Kerr nonlinearity have been studied and it
has been shown that the refractive Kerr nonlinearity is enhanced
using EIT. Besides, these close systems also have broad range of
applications that stimulated our interest to this system. For
example, they have been considered as perspective candidates for
realization of stop-and-go slow light~\cite{agarwal01pra,
rost02jmo}, backward scattering~\cite{rost06prl,rost07ieee}.
Furthemore, the interest to this topic is stimulated by recent
work~\cite{q-storage} in which a quantum storage based on
electromagnetically induced transparency has been predicted. The
first experiments in support of the theoretical predictions have
also been performed \cite{q-storage}. In \cite{zibrov03prl}, it was
shown that the quantum state of light can be stored and retrieved in
a dense medium by using the different regimes of switching on and
off a control field. A quantum state of light having one
polarization and carrier frequency can be transfered to the same
state of light but having a different carrier frequency,
polarization, or direction of propagation. These systems with
microwave field have better controlled probe transparency because
the absorption of the microwave field is much smaller than optical
fields, which is important for improving and optimizing quantum
storage efficiency~\cite{ira-storage-prl, storage-qphys}. Slow light
produces delay that can be used in optical buffers, the delay time
is limitted by the absorption of probe field. Using an auxiliary
microwave field can improve the important parameter for broadband
systems, the product of delay time and bandwidth of the
pulse~\cite{qing05pra}, which shows the effective number of
communication channels.

In this paper, we report the study of EIT in a three-level $\Lambda$ scheme
interacting with two optical fields and a microwave
field coupled with hyperfine levels. ``Perturbing''
(due to a microwave field) the coherence of two ground states leads
the change of maximum transmission of probe field. Either
constructive or destructive EIT peak can be obtained depending on the
relative phase between the optical fields and the microwave field. The paper
is organized by starting with a simple theoretical model of
a close-$\Lambda$ scheme. Then, we describe experimental details and obtained
results. At the end we present numerical simulations that reproduce
experimental results.

\section{Theory}

\begin{figure}[htb]
\includegraphics[width=0.5\columnwidth]{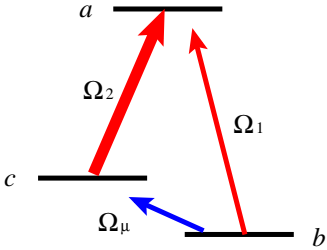}
\caption{\label{fig1theory}(Color online) Energy levels of a closed $\Lambda$
scheme three-level system.}
\end{figure}

Let us consider a closed lambda scheme shown in Fig.~\ref{fig1theory}, in
which a three-level atomic medium is coupled with two optical fields and a
microwave field between two ground states.

\begin{figure*}[htb]
\includegraphics[width=\textwidth]{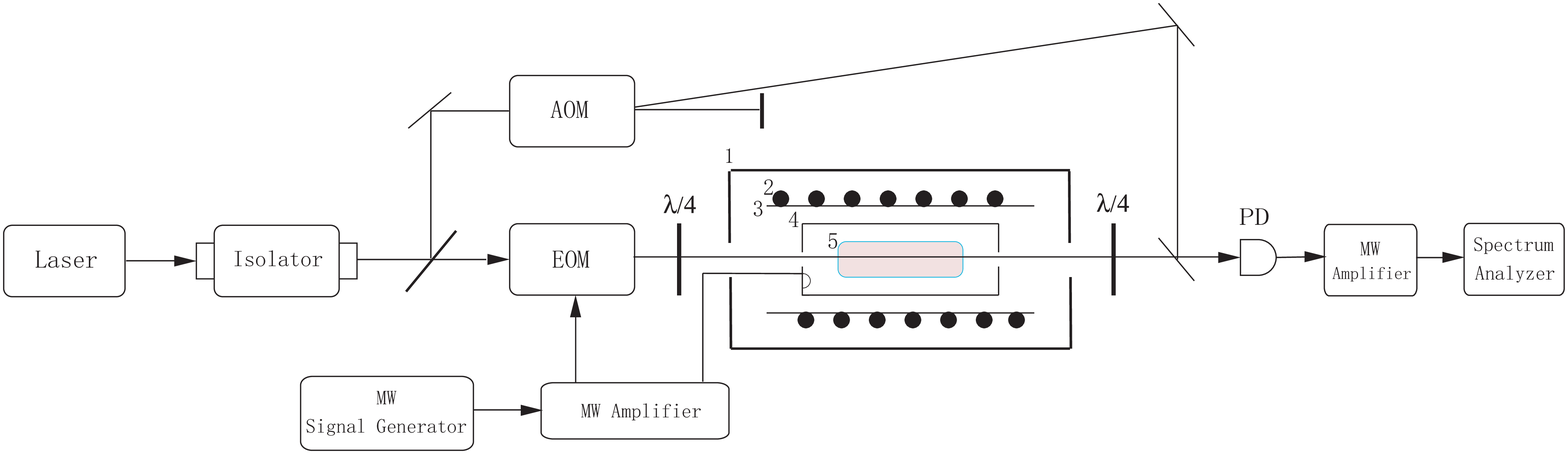}
\caption{\label{fig1}(Color online) Experimental setup. EOM - electro-optic
modulator; AOM - acousto-optic modulator; D - photodiode; the oven
is assembled with 1. copper tube; 2. non-magnetic heater; 3. magnetic
shield; 4. microwave cavity with antenna; 5. Rb cell.}
\end{figure*}

The Hamiltonian of the system can be written as
\begin{eqnarray}
H&=&H_0+H_I;\\
H_0&=&\hbar \omega_a |a\rangle \langle a|+\hbar \omega_b |b\rangle
\langle
b|+\hbar \omega_c |c\rangle \langle c|; \\
H_I&=&-\hbar [\Omega_1 e^{-i \nu_1 t}|a\rangle \langle b|+\Omega_2
e^{-i \nu_2 t}|a\rangle \langle c|\nonumber \\&&+\Omega_{\mu} e^{-i
\nu_\mu t}|c\rangle \langle b|+h.c.];
\end{eqnarray}
where $\Omega_1$, $\Omega_2$ and $\Omega_{\mu}$ are rabi frequencies
of the optical probe field, the optical driving field and the
microwave field respectively; $\nu_1$, $\nu_2$ and $\nu_{\mu}$ are
angular frequencies of corresponding fields. The density matrix
equation of motion is given by
\begin{equation}
\dot{\rho}=-\frac{i}{\hbar} [H,\rho] - \frac{1}{2} \{ \Gamma, \rho
\}
\end{equation}
where $\{ \Gamma, \rho \}=\Gamma \rho+\rho \Gamma$, and $\Gamma$ is
the relaxation matrix. The non-diagonal elements of the density
matrix equations are found as the following,

\begin{eqnarray}
\dot{\rho}_{ab}&=&- \Gamma_{ab} \rho_{ab} - i
\Omega_1(\rho_{aa}-\rho_{bb})+i\Omega_2 \rho_{cb} \nonumber \\&&-
i\Omega_\mu e^{i(\nu_1-\nu_2-\nu_\mu)t} \rho_{ac} \\
\dot{\rho}_{ac}&=&- \Gamma_{ac} \rho_{ac} - i \Omega_2
(\rho_{aa}-\rho_{cc}) + i \Omega_1 \rho_{bc} \nonumber \\&& - i
\Omega_{\mu}^* e^{-i(\nu_1-\nu_2-\nu_\mu)t} \rho_{ab}\\
\dot{\rho}_{cb}&=&-\Gamma_{cb} \rho_{cb} - i \Omega_\mu
e^{i(\nu_1-\nu_2-\nu_\mu)t} (\rho_{cc}-\rho_{bb}) \nonumber \\&&+ i
\Omega_2^* \rho_{ab} - i \Omega_1 \rho_{ca}
\end{eqnarray}
where $\Gamma_{ab}=\gamma_{ab}+i(\omega_{ab}-\nu_1)$,
$\Gamma_{ac}=\gamma_{ac}+i(\omega_{ac}-\nu_2)$ and
$\Gamma_{cb}=\gamma_{cb}+i(\omega_{cb}+\nu_2-\nu_1)$. We consider
the case in which the driving field is on
resonant($\nu_2=\omega_{ac}$), while the probe field and microwave
field have the same detuning $\Delta\equiv\omega_{ab}-\nu_1 =
\omega_{cb}-\nu_\mu  $, thus $\nu_1 - \nu_2 - \nu_\mu=0$. In the
steady-state
regime($\dot{\rho}_{ab}=\dot{\rho}_{ab}=\dot{\rho}_{cb}=0$),
assuming that the driving field is much stronger than the probe
field ($|\Omega_2| \gg |\Omega_1|$) so that almost all of the
population remains in the ground state $|b\rangle$, i.e.
$\rho_{bb}\simeq1$ and $\rho_{aa}=\rho_{cc}\simeq0$, we can solve
equations (5-7) for $\rho_{ab}$,
\begin{equation}
\rho_{ab}=\frac{i
\Gamma_{cb}\Omega_1}{\Gamma_{ab}\Gamma_{cb}+|\Omega_2|^2}-\frac{\Omega_2
\Omega_\mu}{\Gamma_{ab}\Gamma_{cb}+|\Omega_2|^2}
\end{equation}
with $\Gamma_{ab}=\gamma_{ab}+i\Delta$ and
$\Gamma_{cb}=\gamma_{cb}+i\Delta$. The propagation equation of probe
field is give by
\begin{equation}
\frac{\partial\Omega_1}{\partial z}+ik_1\Omega_1=-i\eta\rho_{ab};
\end{equation}
where $\eta=\nu_1 N \wp_{ab}^2/(2\epsilon_0 c \hbar)$ is the
coupling constant, $N$ is the atomic density, $\wp_{ab}^2$ is the
dipole moment of the transition $|a\rangle \leftrightarrow
|b\rangle$, $\epsilon_0$ is the permittivity in vacuum. Consider
optical fields as plane waves:
\begin{equation}
\Omega_i(z,t)=\tilde{\Omega}_i(z,t) e^{-ik_iz},
\end{equation}
where $\tilde{\Omega}_i$($i=1,2$) are slowly varying amplitudes in
the envelopes of optical fields in space, and $k_i$($i=1,2$) are
wave numbers of optical fields. With these expressions and equation
(8), the propagation equation of probe field can be written as
\begin{equation}
\frac{\partial\tilde{\Omega}_1}{\partial z}=-\eta \frac{\Gamma_{cb}
\tilde{\Omega}_1}{\Gamma_{cb} \Gamma_{ab} + |\tilde{\Omega}_2|^2} -
i \eta \frac{\Omega_\mu\tilde{\Omega}_2 e^{i\triangle k
z}}{\Gamma_{cb} \Gamma_{ab} + |\tilde{\Omega}_2|^2}
\end{equation}
where $\triangle k=k_1 - k_2$.

On the right hand side of equation (11), the first term is due to
the standard Lambda scheme EIT, and the second term is the
contribution from the microwave field. The transmission of probe field
is determined by the interference of these two terms. The second
term is interesting because of the strong dependence on the relative
phase of optical fields and microwave field. This gives us several
ways to control coherence and the transmission of the probe field.
For instant, one can use microwave phase shifter to change the phase
of microwave field; one can also use optical delay line, like the
one used in Ref\cite{Hemmer1990}, to change the phase of optical
field. An alternative way is simply changing the position of the Rb
cell, which is described as the following.

Assume that the driving and probe fields are phase-locked, they form a
wave package along propagation direction with the frequency which is
the frequency difference of two fields. For $^{87}$Rb, this
frequency is 6.835 GHz, and corresponding wavelength is about 4.4
cm. If we put the Rb cell in a microwave cavity which is excited by
a microwave with frequency 6.835 GHz, the phase of the microwave in
a cavity does not change when we move the cell and the microwave cavity
together. However, the relative optical phase changes since the
relative position of the cell with respect to the wave package of
optical fields changes. In other words, we are able to change
the phase $\triangle k z$ by moving the cell and microwave cavity
along the propagation direction of optical fields.

\section{Experiment}
\subsection{Experimental setup}

\ The experimental setup is schematically shown in Fig. 2. A diode
laser is tuned to the D$_1$ resonance line of $^{87}$Rb atoms,
specifically at the $5S_{1/2}(F=2)\leftrightarrow 5P_{1/2}(F=2)$
transition. The laser beam passes through an electro-optic modulator
(EOM) which is driven by a microwave with frequency 6.835 GHz, and
two sidebands are generated. One of them is working as probe field
at the $5S_{1/2}(F=1)\leftrightarrow 5P_{1/2}(F=2)$ transition.
Another sideband is 6.835 GHz frequency downshifted with respect to
the drive field, this downshifted field is far from resonance and
has negligible effect on experiment. Another beam is shifted 200 MHz
by an acousto-optic modulator (AOM).

\ The output laser beam from EOM is circularly polarized by a
quarter wave plate, and is directed into a glass cell with the
length of 25 mm. The cell is filled with $^{87}Rb$ vapor and 5 Torr
of Neon buffer gas. The cell is installed in a microwave cavity made
of aluminium and copper. The resonant frequency of microwave cavity
is 6.835 GHz, and the loaded quality factor is $Q\approx2000$. The
microwave injected into the cavity comes from the same signal
generator which also provides the driving microwave for EOM. The
microwave cavity with Rb cell is installed in a magnetic shield. A
non-magnetic heater is used to control the temperature.

\ With the optical fields (drive and probe) coming out from EOM and
the microwave field in the cavity, we have a closed-Lambda system as
shown in Fig. 1. During the experiment, the microwave generator is
200 kHz frequency modulating around 6.835 GHz. Therefore, the probe
laser field and the microwave field are synchronized to be 200 kHz
frequency scanning. The transmitted probe field is detected by the
heterodyne detection described in Ref\cite{Kash1999}. The
transmitted light is beating with an additional optical field which
is 200 MHz frequency shifted by an acousto-optic modulator(AOM), two
sidebands(one of them is the probe field) are separated by 400 MHz
in the beating signal which is detected by a fast photo detector
with the bandwidth of 25 GHz. The signal from photo detector is
acquired by a spectrum analyzer which is synchronized with the
modulation of microwave generator, and the center frequency is set
at the hyperfine splitting frequency plus 200 MHz. The amplitude of
beating signal at this frequency is proportional to the transmission
of probe field.

\subsection{Experimental results}

\begin{figure}[htb]
\includegraphics[width=\columnwidth]{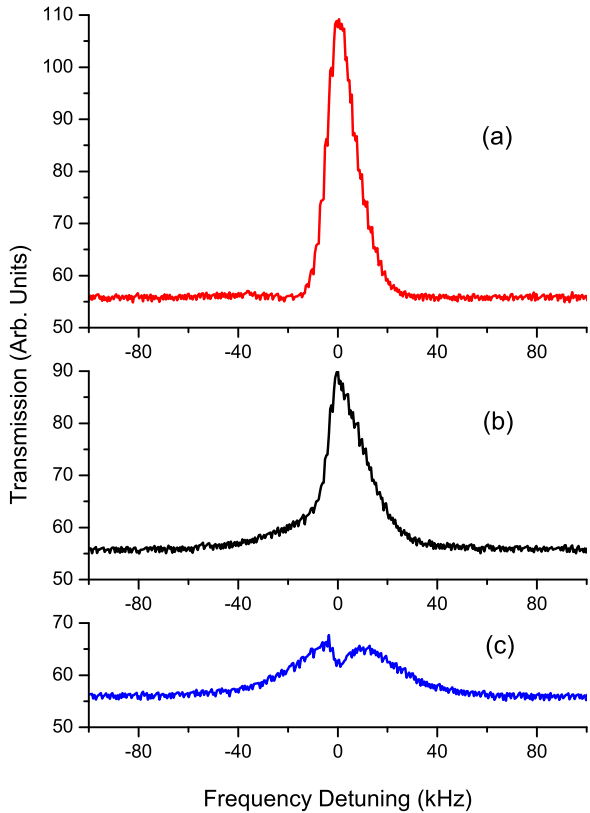}
\caption{\label{fig2}(Color online) Transmitted probe field intensity
recorded by spectrum analyzer. (a) constructive transmission with
microwave field applied; (b) transmission without microwave field
applied; (c) destructive transmission with microwave field applied.}
\end{figure}

\begin{figure*}[htb]
  \centerline{
    \mbox{\includegraphics[width=0.4\textwidth]{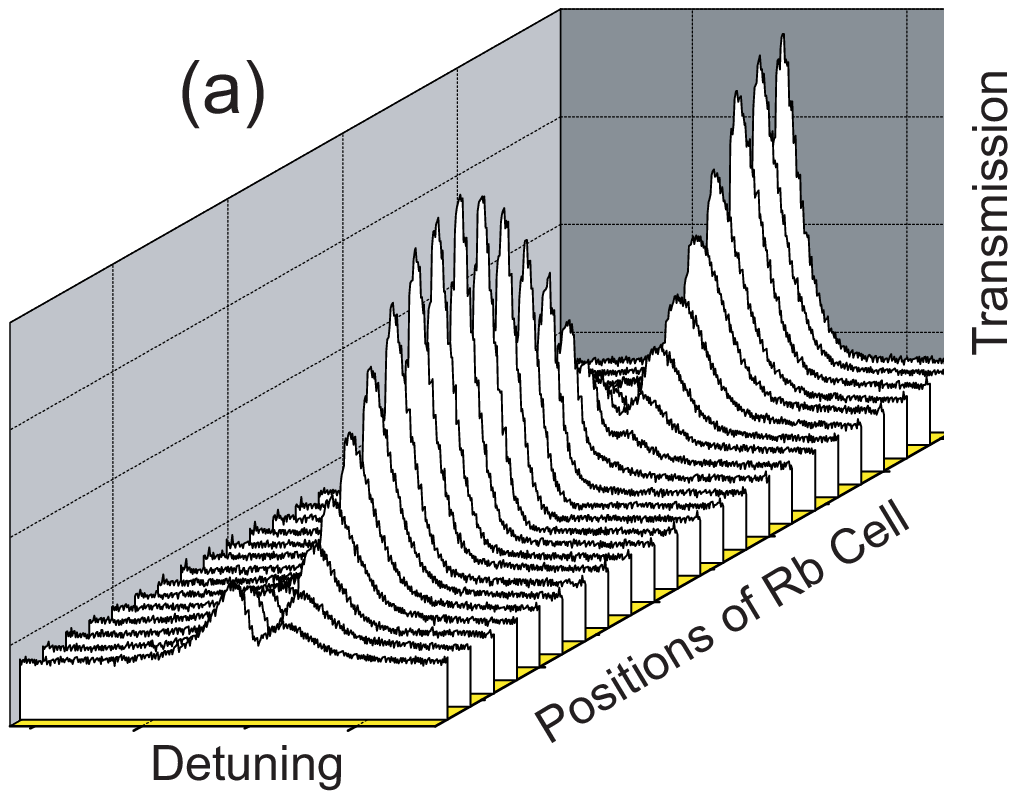}}
    \mbox{\includegraphics[width=0.4\textwidth]{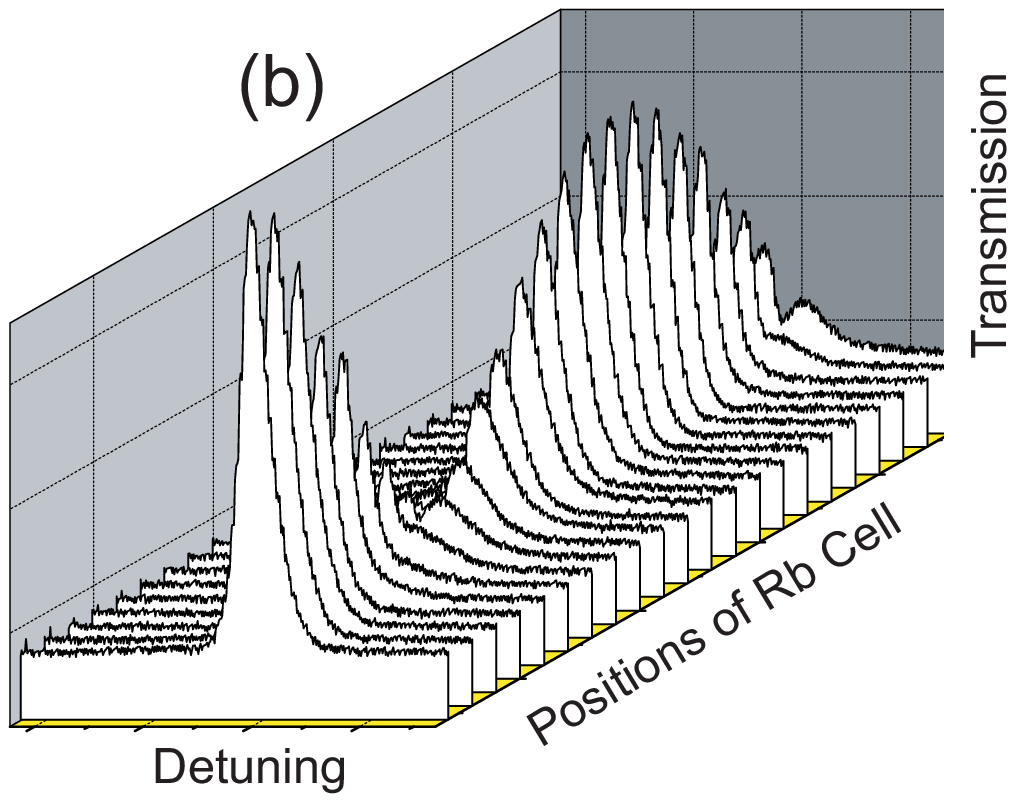}}
  }
  \caption{(Color online) The EIT peaks as we change the position of cell along the optical fields. (a) and (b) correspond to the case where the input laser fields are right and left, respectively, circularly polarized. The distance between
two maximums (or minimums) next to each other is about 4.4 cm.}
\end{figure*}

\ Without applying microwave field, as we vary the detuning, the
transmission is varying. The EIT transmission peak is shown in Fig.
3(b). Applying a microwave field changes the transmission of probe
field. As discussed above, we change the relative phase between
optical fields and microwave field by changing the position of cell
and microwave cavity along the optical axis. Due to the interference
of two terms on the right hand side of equation (11), the
transmission of probe field could be either constructive or
destructive depending on the relative phase. As shown in Fig. 3,
both constructive and destructive transmission of probe field have
been observed. Curve (a) and (c) correspond maximum and minimum
transmission of probe field respectively, as we move the Rb cell
(i.e. change the relative phase).

\ An interesting feature needs to be pointed out for the case of
destructive transmission (Fig. 3c). In this case, the amplitude of
EIT peak decreases as we expected, and we also have a small dip on
the top which indicates that one (the one due to presence of
microwave field) of interfering terms has relatively narrower width.
Its width is narrower than EIT width.

\ The EIT peak is recorded at every 3 mm we move the cell along the
optical axis. Fig. 4 shows how the EIT peak changes as we move the
Rb cell. Fig. 4(a) is the result obtained with right circularly
polarized input laser field, and Fig. 4(b) is the result obtained
with left circularly polarized input laser field. The amplitude of
EIT peak is oscillating with the change of cell position. The
distance between two maximums (or minimums) next to each other is
about 4.4 cm, which is exactly the wavelength of beating envelope of
input optical fields. This periodicity is consistent with the
theoretical prediction described above.

\ The oscillation is clearly shown in Fig. 5, where we plot the
amplitude of EIT peaks as a function of relative phase (phase $2\pi$
corresponds the wavelength 4.4 cm). The dash lines are fittings of
sinusoid function. Comparing the cases of right and left circularly
polarized input laser fields, the behaviors are exactly opposite.
This feature is very surprising, because the whole system is
symmetrical about the optical axis, and there is no obvious way to
tell the difference between left and right circular polarizations.
However, atoms are smart enough to see the difference. Left and
right circularly polarized fields are coupled with different Zeeman
sub-levels, the corresponding magnetic moments have opposite signs
which introduce a phase difference of $\pi$ in our results.

\begin{figure}[htb]
\includegraphics[width=0.9\columnwidth]{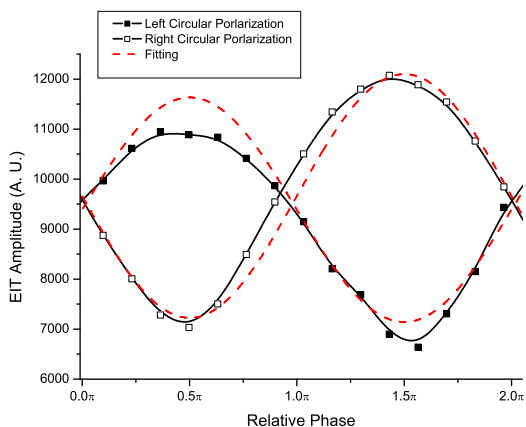}
\caption{\label{fig5}(Color online) The amplitude of EIT peaks dependence
on relative phase. Red and blue curves correspond to the cases of
left and right circularly polarized input laser fields. Dash lines
are fittings of sinusoid function.}
\end{figure}

\begin{figure}[hbt]
\includegraphics[width=0.9\columnwidth]{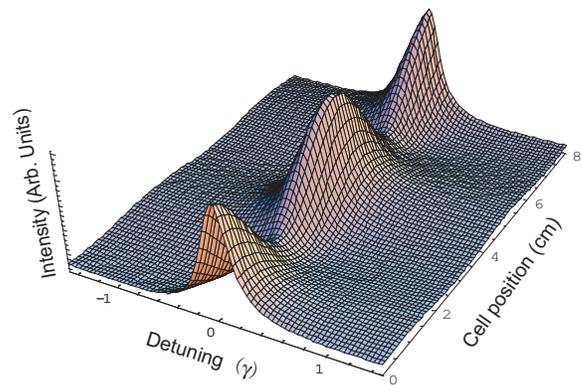}
\caption{\label{fig6}(Color online) Numerical simulation of the
transmission of probe field dependence on detuning and cell
positions. In the simulation, we use $\gamma_{ab}=5$,
$\gamma_{bc}=10^{-3}$, $\Omega_{10}=0.1$, $\Omega_2=1$,
$\Omega_\mu=0.02$, $\eta=0.9$, $L=2.5\ cm$ and $\Delta k=1.5\
cm^{-1}$.}
\end{figure}

\section{Simulation}

To gain physical insights for the obtained results, we
perform simulation based on the equation (11).
Assume that the length of the Rb cell be $L$, and the optical fields
enter the Rb cell at position $z_0$ and leave
at position $z_0+L$. With the probe field
$\tilde{\Omega}_{10}$ entering the cell, equation (11) gives the
transmitted probe field $\tilde{\Omega}_1$ as the following,

\begin{eqnarray}
&&\tilde{\Omega}_1(z_0+L)= \tilde{\Omega}_{10}e^{-\alpha L} -i
\frac{\eta \Omega_{\mu}
\tilde{\Omega}_2}{\Gamma_{cb}\Gamma_{ab}+|\tilde{\Omega}_2|^2}\nonumber
\\&&\frac{1}{i\Delta
k+\alpha} [e^{i\Delta k(z_0+L)}-e^{i\Delta k z_0-\alpha L}];
\end{eqnarray}

where $\alpha$ is the absorption coefficient which is given by
\begin{equation}
\alpha=\eta
\frac{\Gamma_{cb}}{\Gamma_{cb}\Gamma_{ab}+|\tilde{\Omega}_2|^2} \ .
\end{equation}

The simulation result is shown in Fig. 6. The parameters we used in
the simulation are the following: $\gamma_{ab}=5$,
$\gamma_{bc}=10^{-3}$, $\Omega_{10}=0.1$, $\Omega_2=1$,
$\Omega_\mu=0.02$, $\eta=0.9$, $L=2.5\ cm$ and $\Delta k=1.5\
cm^{-1}$. As varying the detuning, the maximum transmission appears
at zero detuning. Meanwhile, the maximum transmission is oscillating
when we change $z_0$ which determines the position of Rb cell, and
the period of oscillation is about 4.4 cm. The simulation shows the
similar behavior as the experimental results, except for the dip at
EIT peaks in destructive cases.

This narrow feature, the dip in the EIT peak, could be used for
EIT-based applications such as improving accuracy of atomic clock.
Eventhough, the simple model described above do not predict this
narrowing. A detail theoretical investigation of this feature should
include four-wave mixing.

It is interesting to note that
the obtained results can be considered
for realization of the stop-and-go slow light~\cite{agarwal01pra, rost02jmo}.
Using the microwave field that has a frequency being close to a resonance,
the dispersion can be modified in a controllable way that slows or accelerates
the group velocity of light.

The obtained results can be also applied to the backward scattering
predicted in \cite{rost06prl,rost07ieee}. By controlling dispersion of the
medium with the optical fields, a microwave field can be produced.
Its direction of generation is determined by the parameters of the fields,
in particular, the detuning of the optical fields from the two-photon
resonance.

Furthermore, the interest to this topic is stimulated by the recent
work~\cite{q-storage}
in which a quantum storage based on electromagnetically induced
transparency has been predicted.
Because absorption of the microwave field is
much smaller than optical fields,
these systems have
better controlled probe transparency, which is important for improving and
optimizing efficiency of quantum
storage~\cite{ira-storage-prl, storage-qphys}.
Slow light produces delay that can be used in optical buffers, the delay time
is limitted by the absorption of probe field. Using auxiliary microwave field
can improve the product of delay time and bandwidth of
the pulse~\cite{qing05pra}.
The broad range of applications stimulated our interest to
the atomic system with the optical and microwave fields.

\section{Conclusion}

In conclusion, we have experimentally studied EIT in Rb atoms
coupled with two optical fields and a microwave field. The microwave
is coupled to two hyperfine levels, and coherently ``perturbs'' the
coherence of two hyperfine levels, thus change the transmission of
probe field. It has been found that the maximum transmission of
probe field depends on the relative phase between optical fields and
microwave field, and both constructive and destructive EIT peaks
have been observed. A simple theoretical model and a numerical
simulation have been provided. The simulation shows the similar
behavior as the experimental results. However, a more detailed
theoretical model is required to explain the dip which occurs in
destructive EIT peaks.

\section{Acknowledgments}

We thank Chang Shin for his help on preparation of microwave cavity.
This work is supported by the Welch Foundation (Grant A-1261), the
NSF grant EEC-0540832 (MIRTHE ERC) and the Office of Naval Research
(N00014-07-1-1084 and N0001408-1-0948).

\bibliography{phase_EIT}

\end{document}